%% file: main.tex
\documentclass[a4paper,11pt]{article}
\pdfoutput=1

\usepackage{jheppub}

\usepackage[usenames,dvipsnames]{xcolor}
\usepackage[english]{babel}
\usepackage{enumerate,enumitem,float,graphicx,comment}

\usepackage[T1]{fontenc}
\usepackage[utf8]{inputenc}
\usepackage{lmodern}
\usepackage[normalem]{ulem}

\makeatletter
\g@addto@macro\bfseries{\boldmath}
\makeatother

\usepackage{amsmath,amssymb,mathtools}

\usepackage{tabulary}
\newcolumntype{K}[1]{>{\centering\arraybackslash}p{#1}}

\input{defs.tex}

\newcommand{\eq}[1]{Eq.(\ref{#1})}

\def\be{\begin{equation}}
\def\ee{\end{equation}}
\def\ba{\begin{eqnarray}}
\def\ea{\end{eqnarray}}
\def\bad{\begin{equation}\begin{aligned}}
\def\ead{\end{aligned}\end{equation}}

\def\red{\color{red}}

\def\td{\tilde}
\def\pp{\partial}

\def\mm{\mathrm}
\def\mc{\mathcal}

\def\red{\color{black}}

\title{%
The Weyl double copy in {\red vacuum spacetimes with a cosmological constant}}

\author{Shanzhong Han}
\emailAdd{shanzhong.han@nbi.ku.dk}

\affiliation{%
  The Niels Bohr Institute,
  University of Copenhagen,
  \\
  Blegdamsvej 17,
  DK-2100 Copenhagen Ø,
  Denmark
}

\abstract{
{\red We examine} the Weyl double copy {\red relation} for vacuum solutions {\red of the Einstein equations} with a cosmological constant using the {\red approach} we previously {\red described}, in which the spin-$1/2$ massless {\red free-field} spinors (Dirac-Weyl fields) are regarded as basic units. {\red Based} on the exact non-twisting vacuum type N and vacuum type D {\red solutions}, the {\red finding} explicitly {\red shows} that the single and zeroth cop{\red ies fulfill} conformally invariant field equations in conformally flat spacetime. {\red In addition}, irrespective of the presence of a cosmological constant, we {\red demonstrate} that the zeroth copy connects {\red Dirac-Weyl fields} with the degenerate electromagnetic fields in the curved spacetime {\red in addition to connecting} gravity fields with the single copy {\red in conformally flat spacetime}. Moreover, the study {\red also demonstrates the critical significance} the zeroth copy plays in time-dependent radiation {\red solutions}. In particular, for Robinson-Trautman ($\Lambda$) gravitational waves, {\red unlike} the single copy, we find that the zeroth copy carries {\red additional} information to {\red specify} whether the sources of associated gravitational waves are time-like, null, or space-like, at least in the {\red weak} field limit.}

\begin{document}
\maketitle
\flushbottom

\section{Introduction}%
\label{sec:intro}

The double copy originates from the study of perturbative scattering amplitudes\cite{Bern:2008qj,Bern:2010ue,Bern:2019prr}, which brings forth a fascinating connection between gauge amplitudes and gravity amplitudes. Moreover, this idea has been extended to {\red the} classical context. In Kerr-Schild coordinate system,  a map between gravity {\red theory} and gauge theory was proposed, called Kerr-Schild double copy \cite{Monteiro:2014cda}. A wide array of such classes of spacetimes {\red has} been studied \cite{Didenko:2008va,Didenko:2009td,Berman:2018hwd,Luna:2015paa,Ridgway:2015fdl,Luna:2016due,White:2016jzc,Adamo:2017nia,DeSmet:2017rve,Bahjat-Abbas:2017htu,Carrillo-Gonzalez:2017iyj,Ilderton:2018lsf,Lee:2018gxc,Gurses:2018ckx,Lescano:2020nve}. Inspired by this, a new type of double copy {\red relation} called Weyl double copy is drawing more attention \cite{Luna:2018dpt,Keeler:2020rcv,White:2020sfn,Chacon:2021wbr,Chacon:2021hfe,Godazgar:2020zbv,Adamo:2021dfg,Godazgar:2021iae,Easson:2021asd}. {\red This prescription} is represented by
\be\label{sec1:double copy}
\Psi_{ABCD}=\frac{\Phi_{(AB}\Phi_{CD)}}{S},
\ee
where $\Psi_{ABCD}$ is a Weyl spinor describing vacuum gravity fields, $\Phi_{AB}$ is an electromagnetic spinor {\red referring to a Maxwell field in Minkowski spacetime}---the simplest solution of {\red the} gauge theory, and $S$ is an auxiliary scalar field {\red satisfying the wave equation in Minkowski spacetime.} The {\red last} two fields are called single copy and zeroth copy, respectively{\red .} Starting from the gravity fields, the Weyl double copy {\red relation} leads to a gauge field that is {\red completely} independent of the gravity {\red theory}.  {\red As a result, it is thought that}, the Weyl double copy {\red relation could serve} as a {\red link between} gravity theory and gauge theory.

{\red Luna et al. proposed for the first time the Weyl double copy relation for the case of vacuum type D solutions \cite{Luna:2018dpt}.} Then, in spinor language, th{\red is} relation was extended to non-twisting vacuum type N solutions by Godazgar et\ al. \cite{Godazgar:2020zbv}. Making use of the peeling property \cite{Newman:1961qr,Wald:1984rg} of the Weyl tensor, they further showed that the Weyl double copy {\red relation} also holds asymptotically for algebraically general solutions \cite{Godazgar:2021iae}. In addition, at the linearised lever, the Weyl double copy {\red relation} was shown to hold for arbitrary Petrov type solutions using the twistor formalism \cite{White:2020sfn,Chacon:2021wbr}. An extended Weyl double copy {\red prescription} was also proposed recently for non-vacuum solutions{\red ,} whose Weyl spinor {\red is} decomposed into a sum of source terms \cite{Easson:2021asd}. {\red Very recently, r}egarding the Dirac-Weyl (DW) spinors (spin-$1/2$ massless free-field spinors) as the basic units of other higher spin massless {\red free-field} spinors, we systematically {\red revisited} the Weyl double copy {\red relation} for non-twisting vacuum type N and vacuum type D solutions \cite{Han:2022ubu}. {\red We further} found a map similar to the {\red Weyl} double copy {\red prescription} for non-twisting vacuum type III spacetimes. 

However, the Weyl double copy {\red relation} for the exact vacuum solutions with a cosmological constant has not {\red yet} been {\red investigated}. This is the {\red primary objective} of the {\red current effort}.  In fact, since 1998, by the observations of supernovae of Ia type \cite{SupernovaSearchTeam:1998fmf,SupernovaCosmologyProject:1998vns}, studies {\red have shown} that the expansion of our universe is accelerating, which strongly supports the condition that the cosmological constant $\Lambda$ is nonzero and positive. On the other hand, although Anti-de Sitter (AdS) spacetime does not appear to have direct cosmological applications, it plays a {\red crucial} role in AdS/CFT correspondence. Therefore, {\red investigating the Weyl double copy relation in the presence of a cosmological constant would be of interest. Currently, there are two possible research directions: one is to interpret the cosmological constant as a source of the single and zeroth copies in the flat spacetime; the other is to consider the (A)dS spacetime to be the background of the single and zeroth copies. The former idea was proposed for the first time in Kerr-Schild double copy in Taub-NUT spacetime \cite{Luna:2015paa} and it would be natural in the direct investigation of the relationship between gravity theory and gauge theory. On the other hand, the latter can be viewed as a precursor to the former. Moreover, it is also advantageous for extending the remit of the Weyl double copy, including cosmological applications and perturbation theory. This has been done in Ref. \cite{Carrillo-Gonzalez:2017iyj} for} Kerr-Schild($\Lambda$) double copy, {\red which shows that} the single and zeroth copies satisfy different equations for time-dependent and time-independent solutions. {\red These outcomes} encourage us to {\red study} whether {\red or not} the Weyl double copy {\red relation} shares this property. In this paper, we shall give an explicit demonstration to show {\red that,} different from the Kerr-Schild($\Lambda$) double copy, the single and zeroth copies {\red in the} Weyl double copy {\red prescription} all satisfy conformal invariant field equations {\red in} conformally flat spacetime, both for time-independent solutions and time-dependent solutions. {\red Our finding} coincides with the statement of Ref. \cite{White:2020sfn} in the twistorial version. Some {\red interesting} relations between the zeroth copy and gravitational waves will also be discussed.

The structure of this paper is as follows. In Sec. \ref{sec2}, we will briefly review how to construct electromagnetic spinors in vacuum type N and type D spacetimes by regarding DW spinors as the basic units . Then, we will study the Weyl double copy {\red for} exact vacuum solutions with a cosmological constant in Sec. \ref{sec3}. The interpretations of the single copy and the zeroth copy will also be included. Discussion and conclusions are given in Sec. \ref{sec4}. The notation of this paper follows the conventions of Ref. \cite{Han:2022ubu}.

\section{Massless free-fields in spinor formalism}
\label{sec2}
In this section, we will briefly review how to construct electromagnetic spinors in order to verify the Weyl double copy {\red relation} {\red using the methodology} of {\red the} previous work \cite{Han:2022ubu}.

In spinor formalism, spin-${\red k}/2$ massless free-field equations have a simple form \cite{Penrose:1985bww}
\be\label{sec2:general massless free-field equation}
\nabla^{A_1A'_1}\mc{S}_{A_1A_2...A_{\red k}}=0,
\ee
where the spinor $\mc{S}_{A_1 A_2...A_{\red k}}$ is totally symmetric. 

For spin-$2$ massless free-field{\red s}, the {\red spinor $\mc{S}$ refers to} the Weyl {\red spinor $\Psi_{ABCD}$ translated from the Weyl }tensor $C_{abcd}$ 
\be\label{sec2: Weyl spinor}
C_{abcd}=C_{AA'BB'CC'DD'}=\Psi_{ABCD}\varepsilon_{A'B'}\varepsilon_{C'D'}+\bar{\Psi}_{A'B'C'D'}\varepsilon_{AB}\varepsilon_{CD}.
\ee
It is easy to find that the Weyl spinor $\Psi_{ABCD}$ plays the same role as the Weyl tensor $C_{abcd}$. For a vacuum spacetime (with or without a cosmological constant $\Lambda$), the Einstein field equation is absorbed into the Bianchi identity, {\red which reads}
\be\label{sec2:4eom}
\nabla^{AA'}\Psi_{ABCD}=0.
\ee 
This is nothing but a spin-$2$ massless free-field equation. Notably, the fact that this field equation {\red remains} the same {\red regardless of the presence or absence of} a cosmological constant {\red motivates} us to generalize the original Weyl double copy to the case with a cosmological constant. As is well known, ten independent real components of the Weyl tensor {\red can be reduced to} 5 independent complex scalars with the aid of a null tetrad, as defined in Ref.\cite{Chandrasekhar:1985kt}. Using the totally symmetric property of the Weyl spinor, we can define them as follows,
\be
\begin{aligned}
\psi_0 &=\Psi_{ABCD}o^A o^B o^C o^D=C_{abcd}\ell^a m^b \ell^c m^d ,\\
\psi_1 &=\Psi_{ABCD}o^A o^B o^C \iota^D=C_{abcd}\ell^a m^b \ell^c n^d ,\\
\psi_2 &=\Psi_{ABCD}o^A o^B \iota^C \iota^D=C_{abcd}\ell^a m^b \bar{m}^c n^d ,\\
\psi_3 &=\Psi_{ABCD}o^A \iota^B \iota^C \iota^D=C_{abcd}\ell^a n^b \bar{m}^c n^d ,\\
\psi_4 &=\Psi_{ABCD}\iota^A \iota^B \iota^C \iota^D=C_{abcd}\bar{m}^a n^b \bar{m}^c n^d ,
\end{aligned}
\ee
where the second equations hold based on the definition of the null tetrad in the spinor bases
\be\label{sec2:NP}
\begin{array}{llll}
\ell^{a}=o^{A} \bar{o}^{A^{\prime}}, & n^{a}=\iota^{A} \bar{\iota}^{A^{\prime}}, & m^{a}=o^{A} \bar{\iota}^{A^{\prime}}, & \bar{m}^a=\iota^A\bar{o}^{A^{\prime}} ,\\
\ell_{a}=o_{A} \bar{o}_{A^{\prime}}, & n_{a}=\iota_{A} \bar{\iota}_{A^{\prime}}, & m_{a}=o_{A} \bar{\iota}_{A^{\prime}},&\bar{m}_a=\iota^A\bar{o}^{A^{\prime}}.
\end{array}
\ee
It is easy to check that the above correspondence indeed defines a null tetrad such that 
\be
\begin{aligned}
\ell^2=n^2=m^2=\bar{m}^2=0,\\
 \ell \cdot n=1,\quad m\cdot \bar{m}=-1,\quad \ell\cdot m=n\cdot m=\ell\cdot \bar{m}=n\cdot \bar{m}=0.
 \end{aligned}
\ee
The spin coefficients are defined the same as in the preceding work \cite{Han:2022ubu}{\red , 
\be
\begin{aligned}
&\kappa^{\red *}=m^{\bold{a}}\ell^{\bold{b}}\nabla_{\bold{b}}\ell_{\bold{a}}, \quad &\pi^{\red *}=n^{\bold{a}}\ell^{\bold{b}}\nabla_{\bold{b}}\bar{m}_{\bold{a}}, \quad &\epsilon^{\red *}=\frac{1}{2}(n^{\bold{a}}\ell^{\bold{b}}\nabla_{\bold{b}}\ell_{\bold{a}}+m^{\bold{a}}\ell^{\bold{b}}\nabla_{\bold{b}}\bar{m}_{\bold{a}}),\\
&\tau^{\red *}=m^{\bold{a}}n^{\bold{b}}\nabla_{\bold{b}}\ell_{\bold{a}}, \quad &\nu^{\red *}=n^{\bold{a}}n^{\bold{b}}\nabla_{\bold{b}}\bar{m}_{\bold{a}}, \quad &\gamma^{\red *}=\frac{1}{2}(n^{\bold{a}}n^{\bold{b}}\nabla_{\bold{b}}\ell_{\bold{a}}+m^{\bold{a}}n^{\bold{b}}\nabla_{\bold{b}}\bar{m}_{\bold{a}}),\\
&\sigma^{\red *}=m^{\bold{a}}m^{\bold{b}}\nabla_{\bold{\red{b}}}\ell_{\bold{{\red a}}},\quad &\mu^{\red *}=n^{\bold{a}}m^{\bold{b}}\nabla_{\bold{b}}\bar{m}_{\bold{a}}, \quad &\beta^{\red *}=\frac{1}{2}(n^{\bold{a}}m^{\bold{b}}\nabla_{\bold{b}}\ell_{\bold{a}}+m^{\bold{a}}m^{\bold{b}}\nabla_{\bold{b}}\bar{m}_{\bold{a}}),\\
&\rho^{\red *}=m^{\bold{a}}\bar{m}^{\bold{b}}\nabla_{\bold{b}}\ell_{\bold{a}}, \quad &\lambda^{\red *}=n^{\bold{a}}\bar{m}^{\bold{b}}\nabla_{\bold{b}}\bar{m}_{\bold{a}}, \quad &\alpha^{\red *}=\frac{1}{2}(n^{\bold{a}}\bar{m}^{\bold{b}}\nabla_{\bold{b}}\ell_{\bold{a}}+m^{\bold{a}}\bar{m}^{\bold{b}}\nabla_{\bold{b}}\bar{m}_{\bold{a}}).
\end{aligned}
\ee}For more details, {\red one may refer to} Ref. \cite{Stewart:1990uf, Chandrasekhar:1985kt}. To distinguish from other symbols in this paper, we use $*$ to mark these spin coefficients in the following, such as $\kappa^*$, $\alpha^*$, $\beta^*$, etc. 
Expanding out the Weyl spinor, the general form reads
\be\label{sec2:general Weyl}
\begin{aligned}
\Psi_{ABCD} &=\psi_0 \iota_A\iota_B\iota_C\iota_D-4\psi_1o_{(A}\iota_B\iota_C\iota_{D)}+6\psi_2o_{(A}o_B\iota_C\iota_{D)}\\
&-4\psi_3o_{(A}o_Bo_C\iota_{D)}+\psi_4o_{A}o_Bo_C o_{D}.
\end{aligned}
\ee
For vacuum type N and type D solutions, the Weyl spinors {\red are reduced} to
\begin{align}
\text{type\ N}:\quad \Psi_{ABCD}&=\psi_4o_{A}o_Bo_C o_{D}\label{sec2:N4},\\
\text{type\ D}:\quad \Psi_{ABCD}&=6\psi_2o_{(A}o_B\iota_C\iota_{D)}\label{sec2:D4}.
\end{align}

For spin-$1/2$ massless free-field{\red s, the spinor $\mc{S}$ refers to a DW spinor $\xi_A$}. {\red \eq{sec2:general massless free-field equation} in this case represents the DW field equation}
\be\label{sec2:DW equation}
\nabla^{AA'}\xi_A=0.
\ee
With the map proposed in the preceding work \cite{Han:2022ubu}
\be\label{sec2:14map}
\Psi_{ABCD}=\frac{\xi_{(A}\eta_B\zeta_C\chi_{D)}}{S_{14}}, 
\ee
where the four DW spinors on the right side can be chosen to be the same (depending on which type of spacetime we are focusing on), we are {\red now able} to derive the DW spinors in a {\red certain} vacuum spacetime with a cosmological constant. Correspondingly, the electromagnetic spinors in curved spacetime will be formulated. 

Specifically, for vacuum type N solutions, the map \eq{sec2:14map} reduces to
\be\label{sec2:N14}
\Psi_{ABCD}=\frac{\xi_{(A}\xi_B\xi_C\xi_{D)}}{S_{14}}=\frac{\xi_{(A}\xi_B\xi_C\xi_{D)}}{(S_{12})^3}.
\ee
 From \eq{sec2:N4} one can see that {\red $\xi_A = \xi o_A$}. {\red According to Ref. \cite{Han:2022ubu}, we know that $S_{12}=S_{24}=S_{14}^{1/3}$, where $S_{ij}$ is an auxiliary scalar connecting a spin-$i/2$ massless free-field spinor with a spin-$j/2$ massless free-field spinor.}
 The independent dyad components of the Weyl field equation \eq{sec2:4eom} then read
\begin{align}\label{sec2: Weyl field equation}
o_A\nabla^{AA'}\log\ \Psi_4+4o_A \iota^B \nabla^{AA'}o_B-\iota_A o^B \nabla^{AA'}o_B=0,
\end{align}
where we define the Weyl scalar $\Psi_4=\psi_4$. {\red On the other hand, t}he dyad component of the DW field equation {\red \eq{sec2:DW equation} is given by}
\be\label{sec2:N1}
o_A \nabla^{AA'}\log\ \xi+o_A \iota^B\nabla^{AA'}o_A-\iota_A o^B \nabla^{AA'}o_B=0.
\ee
{\red Combining \eq{sec2:N14}, \eq{sec2: Weyl field equation} and \eq{sec2:N1},} the auxiliary scalar $S_{12}$ and the DW scalar {\red $\xi$} will be identified by solving
\be\label{sec3:N12tensor}
\ell\cdot\nabla \log\ S_{12}-\rho^*=0,\quad m\cdot \nabla \log\ S_{12} - \tau^*=0.
\ee
Since there is only one type of DW spinor $\xi_A=\xi o_A$, correspondingly{\red ,} only one type of electromagnetic spinor {\red can exist---}the degenerate electromagnetic spinor
\be\label{sec2:degenerate12}
\Phi_{AB}=\frac{\xi_A\xi_B}{S_{12}}=\frac{\xi^2}{S_{12}} o_Ao_B=\phi_2 o_A o_B.
\ee
{\red Furthermore,} the electromagnetic tensor $F_{ab}=F_{AA'BB'}=\Phi_{AB}\varepsilon_{A'B'}+\bar{\Phi}_{A'B'}\varepsilon_{AB}$, where $\varepsilon_{AB}=2o_{[A}\iota_{B]}$, in the null tetrad we have
\be\label{sec2:dege-tensor}
F_{ab}=2\phi_2 \ell_{[a}m_{b]}+2\bar{\phi}_2\ell_{[a}\bar{m}_{b]}.
\ee

For vacuum type D solutions, {\red many} of spacetimes {\red that} we are familiar with belong to this class, such as Kerr (A)dS black holes, {\red Reissner–Nordström} (A)dS black holes, NUT solutions, $C$-metric, etc. In this case, the map \eq{sec2:14map} reduces to
\be\label{sec2:D14}
\Psi_{ABCD}=\frac{\xi_{(A}\xi_B\eta_C\eta_{D)}}{S_{14}},
\ee
where we choose two DW spinors with the same coefficient, in other words, 
\be\label{sec2:Dxieta}
\xi_A=\xi o_A,\quad \quad \eta_A=\xi \iota_A.
\ee
The dyad components of gravity field equation \eq{sec2:4eom} are then given by
\begin{align}
o_A \nabla^{AA'} \log (\Psi_2) - 3\iota_A o^B\nabla^{AA'}o_B=0,  \label{sec2:D4left}\\
\iota_A \nabla^{AA'} \log (\Psi_2) + 3o_A \iota^B\nabla^{AA'}\iota_B=0, \label{sec2:D4right}
\end{align}
where we let the Weyl scalar $\Psi_2=6\psi_2$. Two dyad components of the DW field equations read
\begin{align}
o_A \nabla^{AA'}\log\ \xi-\iota_A o^B \nabla^{AA'}o_B+o_A \iota^B\nabla^{AA'}o_B=0, \label{sec3:D1left}\\
\iota_A \nabla^{AA'}\log\ \xi+o_A \iota^B\nabla^{AA'}\iota_B-\iota_A o^B \nabla^{AA'}\iota_B=0. \label{sec3:D1right}
\end{align}
By making use of the map \eq{sec2:D14}, the auxiliary scalar $S_{14}$ and the DW scalar will be identified by solving
\be
\begin{aligned}\label{sec2:DS14with4eqs}
\ell\cdot \nabla\log S_{14}+4\epsilon^*-\rho^*=0,\quad m\cdot\nabla\log S_{14}+4\beta^*-\tau^*=0,\\
\bar{m}\cdot\nabla\log S_{14}-4\alpha^*+\pi^*=0,\quad n\cdot \nabla\log S_{14}-4\gamma^*+\mu^*=0.
\end{aligned}
\ee
Different from the type N case, since there are two different types of DW spinors, we thereby have two different types of electromagnetic spinors. Apart from the degenerate electromagnetic spinor we discussed above, the other type is a non-degenerate electromagnetic spinor,
\be\label{sec2:nondegenerate12}
\Phi^{(1)}_{AB}=\phi_1 o_{(A} \iota_{B)}=\frac{\xi^2 o_{(A} \iota_{B)}}{S^{(1)}_{12}}.
\ee
In order to distinguish two different types of electromagnetic spinors, we use a upper index $(1)$ to refer to non-degenerate ones and $(0)$, $(2)$ to refer to degenerate ones $\Phi^{(0)}_{AB}=\phi_0 \iota_A \iota_B$ and $\Phi^{(2)}_{AB}=\phi_2 o_A o_B$, repectively\footnote{{\red The complete expression of the} degenerate electromagnetic spinors {\red in} \eq{sec2:degenerate12} {\red should be} $\Phi^{(2)}_{AB}$, since there is only one type of the field in the type N case, {\red for simplicity, we omit the superscript (2) there.}}. The dyad components of the non-degenerate electromagnetic field equation are given by
\begin{align}
o_A \nabla^{AA'}\log \phi_1 - 2\iota_A o^B \nabla^{AA'}o_B=0 \label{sec3:non-degenerate Maxwell 1},\\
\iota_A \nabla^{AA'}\log \phi_1 + 2o_A \iota^B \nabla^{AA'}\iota_B=0 \label{sec3:non-degenerate Maxwell 2}.
\end{align}
Substitution of the map \eq{sec2:nondegenerate12} into the above equations and multiplying $\bar{o}_{A'}$ and $\bar{\iota}_{A'}$ respectively yield
\be\label{sec2:DS112}
\begin{aligned}
\ell\cdot\nabla\log S^{(1)}_{12}+2\epsilon^*=0,\quad m\cdot \nabla\log S^{(1)}_{12}+2\beta^*=0,\\
\bar{m}\cdot \nabla \log S^{(1)}_{12}-2\alpha^*=0, \quad n\cdot \nabla \log S^{(1)}_{12}-2\gamma^*=0.
\end{aligned}
\ee
Solving the above equations, we are able to obtain the auxiliary scalar field $S^{(1)}_{12}$. The electromagnetic scalar $\phi_1$ will then be determined from \eq{sec2:nondegenerate12}. In analogy to \eq{sec2:dege-tensor}, the non-degenerate electromagnetic tensor in the null tetrad reads
\be\label{sec2:nondege-tensor}
F_{ab}=2\phi_1\left(\ell_{[a}n_{b]}+\bar{m}_{[a}m_{b]}\right)+2\bar{\phi}_1\left(\ell_{[a}n_{b]}+m_{[a}\bar{m}_{b]}\right).
\ee
{\red Correspondingly, the auxiliary scalar field connecting the Weyl field and the non-degenerate electromagnetic field is denoted by $S^{(1,1)}_{24}$, which satisfies
\be
\Psi_2=\frac{(\phi_1)^2}{S^{(1,1)}_{24}},
\ee
where superscript $(1,1)$ corresponds to the product of two electromagnetic scalars $\phi_1$.}

In {\red general}, {\red both for} type N {\red solutions and} type D solutions, once DW spinors are identified, all electromagnetic fields (or other higher spin massless free-fields) {\red in} the curved spacetime principally {\red can} be formulated with the aid of an auxiliary scalar field. To verify the Weyl double copy relation, {\red it is only necessary to locate a specific set of} electromagnetic fields, which {\red are independent of the source parameters or structure functions that determine how the spacetime deviates from the (A)dS background.} If such electromagnetic fields do exist, they {\red will also satisfy the field equation in (A)dS spacetime. These fields are nothing but the} single copy, and the associated auxiliary scalar fields are the zeroth copy.

\section{The Weyl double copy in {\red curved spacetimes}}
\label{sec3}
Regarding DW spinors as basic units, electromagnetic spinors living in a certain curved spacetime are constructed according to {\red the} maps \eq{sec2:degenerate12} and \eq{sec2:nondegenerate12}. Then, they are {\red converted} to tensor form and expanded in terms of the products of the null tetrad bases, such as \eq{sec2:dege-tensor} and \eq{sec2:nondege-tensor}. As we will see later, except for the electromagnetic scalars, the products of the null tetrad bases of the degenerate electromagnetic tensors {\red in non-twisting vacuum type N spacetimes} are independent of {\red the structure functions and source parameters.} {\red The same is true} for non-degenerate electromagnetic tensors {\red in} vacuum type D spacetimes. {\red For the sake of brevity, structure functions and source parameters will be referred to as deviation-information in the following.} Once we {\red have demonstrated} that the electromagnetic scalars are independent of {\red deviation-information, the same will be true of electromagnetic fields. Therefore,} they should satisfy the field equations {\red in} conformally flat spacetime. Surprisingly, one will find that the associated scalar fields automatically satisfy their conformally invariant field equations {\red in} conformally flat spacetime. An explicit demonstration of the Weyl double copy for non-twisting vacuum type N and vacuum type D solutions is given in the following. The signature of the spacetime metric is chosen as $(+,-,-,-)$ in this work.

\subsection{The case {\red of} non-twisting vacuum type N solutions}
As the solutions of gravitational waves, non-twisting vacuum type N solutions ($\Lambda$) are composed of two classes \cite{Bicak:1999ha,Bicak:1999hb}, one is {\red the} non-expanding Kundt($\Lambda$) class, and the other is {\red the} expanding Robinson-Trautman($\Lambda$) class. 
\subsubsection{The Kundt($\Lambda$) class}
The metric in this case reads
\be\label{sec3:Kundt metric}
\mm{d} s^{2}=-F \mm{d} u^{2}+2 \frac{q^{2}}{p^{2}} \mm{d} u \mm{d} v-2 \frac{1}{p^{2}} \mm{d} z \mm{d} \bar{z},
\ee
with
\bad
&p=1+\frac{\Lambda}{6} z \bar{z}, \quad \quad q=\left(1-\frac{\Lambda}{6} z \bar{z}\right) \alpha+\bar{\beta} z+\beta \bar{z}, \\
&F=\kappa \frac{q^{2}}{p^{2}} v^{2}-\frac{(q^{2})_{{\red ,}u}}{p^{2}} v-\frac{q}{p} H, \quad \quad  \kappa=\frac{\Lambda}{3} \alpha^{2}+2 \beta \bar{\beta},\\
&H=H(u, z, \bar{z})=\left(f_{, z}+\bar{f}_{, \bar{z}}\right)-\frac{\Lambda}{3 p}\left(\bar{z} f+z \bar{f}\right).
\ead
where $f$ is an arbitrary complex function of $u$ and {\red $z$}, analytic in {\red $z$}. {\red Further more,} $\alpha$ and $\beta$ are two arbitrary real and complex functions of $u$, respectively. In fact, according to Ref. \cite{Bicak:1999ha}, one can see {\red that} the parameter $\kappa$ is sign invariant. For the case $\Lambda=0$, there are two classes of solutions---generalised pp-waves ($\kappa=0$) and generalised Kundt waves ($\kappa>0$). If our universe admits a positive cosmological constant, namely $\Lambda>0$, there is no limit on $\alpha$ and $\beta$, and there is only one kind of solution---generalised Kundt waves. For the case $\Lambda<0$, the values of parameters $\alpha$ and $\beta$ classify the metric into three types of solutions---generalised Kundt waves ($\kappa>0$), generalised Siklos waves ($\kappa=0$), and generalised pp-waves ($\kappa<0$). We will soon see that the zeroth copy inherits this property {\red to classify} the gravity solutions.

Choosing the null tetrad
\be
\ell=\mm{d} u,\quad n=-\frac{F}{2}\mm{d} u+\frac{q^2}{p^2}\mm{d} v,\quad m=\frac{1}{p}\mm{d}\bar{z},
\ee
we have
\begin{align}
{\red \rho^*=0,\quad \tau^*=-\frac{2\Lambda \bar{z} \alpha+\Lambda\bar{z}^2\beta-6\bar{\beta}}{(6-\Lambda z \bar{z})\alpha+6(z\bar{\beta}+\bar{z}\beta)}},\\
\Psi_4=\frac{1}{72}\left(\Lambda {\red z}\bar{z}+6\right)\left[(\Lambda {\red z}\bar{z}-6)\alpha-6(\bar{z}\beta+z\bar{\beta})\right]\pp^3_{\bar{z}}\bar{f}.
\end{align}
Recalling \eq{sec2:dege-tensor}, it is easy to check 
\be
2\ell_{[a}m_{b]}=\left(\begin{array}{llll}
0\ &\ 0 &\ 0 &\ I \\
0\ &\ 0 &\ 0 &\  0 \\
0\ &\ 0 &\ 0 &\  0 \\
-I\ &\ 0 &\ 0 &\  0 \\
\end{array}\right)\quad \text{and}\quad 
2\ell_{[a}\bar{m}_{b]}=\left(\begin{array}{llll}
0\ &\ 0 &\ I &\ 0 \\
0\ &\ 0 &\ 0 &\ 0 \\
-I\ &\ 0 &\ 0 &\  0 \\
0\ &\ 0 &\ 0 &\  0 \\
\end{array}\right),
\ee
where $I=\frac{6}{6+\Lambda z\bar{z}}$. Both matrices do not depend on the {\red deviation-information}, so {\red the} electromagnetic scalar will decide whether this kind of electromagnetic field is dependent {\red of} the {\red deviation-information} or not. From \eq{sec3:N12tensor}, the auxiliary scalar $S_{12}$ is solved by
\be
S_{12}=\mathcal{C}(u,\bar{z})\frac{\Lambda z\bar{z}+6}{\left(\Lambda z\bar{z}-6\right)\alpha-6\left({\red z\bar{\beta}+\bar{z}\beta}\right)},
\ee
where $\mathcal{C}(u,\bar{z})$ is an arbitrary function of $u$ and $\bar{z}$. Clearly $S_{12}$ itself is independent of the {\red deviation-information}. According to \eq{sec2:N14} and \eq{sec2:degenerate12}, the DW scalar $\xi$ and the electromagnetic scalar $\phi_2$ are solved by
\begin{align}
\xi^4&=\frac{(6+\Lambda z\bar{z})^4{\red \mathcal{C}(u,\bar{z})^3\pp^3_{\bar{z}}\bar{f}}}{72\left[(\Lambda z\bar{z}-6)\alpha-6(\bar{z}\beta+z\bar{\beta})\right]^2},\\
\phi_2&=\frac{(6+\Lambda{\red z}\bar{z})}{6\sqrt{2}}\sqrt{\mathcal{C}(u,\bar{z})\pp^3_{\bar{z}}\bar{f}}\label{sec3:Kundt-phi2}.
\end{align}
The structure function {\red $\bar{f}$}, which measures the value of $\Psi_4$, is absorbed by an arbitrary function $\mathcal{C}(u,\bar{z})$. {\red The e}lectromagnetic scalar thus does not depend on $\pp^3_{\bar{z}}\bar{f}(u,\bar{z})$. So we obtain a {\red particular} degenerate electromagnetic field {\red which is independent of the deviation-information}.  It is easy to check that this type of electromagnetic field satisfies its field equation even for the case $\pp^3_{\bar{z}}\bar{f}=0$. Namely, 
\be
\td{\nabla}_a F^{ab}=0,
\ee
where the symbol tilde denotes that the background is (A)dS spacetimes---conformally flat spacetimes---where we just need to let $f=1$ in the original metric. In fact, there is a freedom to choose a polynomial function $f=c_0(u)+c_1(u)\bar{z}+c_2(u)\bar{z}^2$ as long as $\pp^3_{\bar{z}}\bar{f}(u,\bar{z})=0$, where $c_i(u)$ are expanding parameters of $\bar{z}$. Furthermore, with the fact that the Ricci scalar $R=-4\Lambda$, it is easy to verify that the auxiliary scalar field $S_{24}(=S_{12})$ satisfies the conformally invariant scalar field equation not only {\red in} the curved spacetime but also {\red in} conformally flat spacetime. So we have
\be\label{sec3:0scalar-coformal}
\td{\nabla}^a\td{\nabla}_a S_{24}-\frac{1}{6}\td{R} S_{24}=0.
\ee
When $\Lambda\to0$, the result reduces to the Kundt ($\Lambda=0$) class, the single copy and the zeroth copy satisfy their field equations {\red in} Minkowski spacetime.

More interestingly, one can find that the single copy only confines the structure function. For example, {\red it does not depend on the parameters $\alpha$, $\beta$, and $\Lambda$; the function $f$ in Maxwell scalar {\red only needs to} be a function of coordinates $u$ and {\red $z$}, and there are no other restrictions.} In contrast, the zeroth copy {\red is closely associated} with $\alpha$, $\beta$, and $\Lambda$.  With a negative cosmological constant {\red and in conjunction} with the introduction in the first paragraph of this section, one can see that {\red for different $\kappa$}, it is the zeroth copy that {\red specifies the sort} of curved spacetimes they map.
\subsubsection{The Robinson-Trautman($\Lambda$) class}
One of the familiar {\red form of the} metric {\red for} Robinson-Trautman ($\Lambda$) solutions is given by Garc\'ia D\'iaz and Pleba\'nski \cite{doi:10.1063/1.524843,Bicak:1999ha}
\bad
\mm{d}s^2=-2(A\bar{A}+\psi B)\mm{d}u^2-2\psi \mm{d}u \mm{d}v- 2v\bar{A} \mm{d}u \mm{d}z-2v A \mm{d} u \mm{d}\bar{z} -2v^2 \mm{d}z \mm{d}\bar{z},\\
A=\epsilon z-v f,\quad B=-\epsilon+\frac{v}{2}(f_{,z}+\bar{f}_{,\bar{z}})+\frac{\Lambda}{6}v^2\psi,\quad \psi=1+\epsilon z\bar{z},
\ead
where $\epsilon=+1,0,-1$ corresponds to the source of the transverse gravitational waves {\red being} time-like, null, or space-like, respectively, at least in {\red the weak} field limit. This is consistent with the case {\red that} $\Lambda=0$ \cite{Robinson:1960zzb}. One can also refer to Ref. \cite{Griffiths:2002gm} for more {\red details on the} interpretation {\red of} the Robinson-Trautman solutions. In addition, this metric only depends linearly on an arbitrary structure function $f(u,z)$, {\red which} will help facilitate the following discussions.

Choosing the null tetrad
\be
\ell=du,\quad n=-(A \bar{A}+\psi B)du-\psi dv-\bar{A}v dz-v A d\bar{z},\quad m=vd\bar{z},
\ee
we have
\begin{align}
{\red \rho^*=\frac{1}{v(1+\epsilon z \bar{z})},\quad \tau^*=\frac{\bar{f}}{1+\epsilon z \bar{z}}},\\
\Psi_4=\frac{(1+\epsilon z\bar{z})\pp^3_{\bar{z}}\bar{f}}{2v}.
\end{align}
In this case, the Weyl scalar does not even depend on the cosmological constant. Recalling \eq{sec2:dege-tensor}, one observes
\be
2\ell_{[a}m_{b]}=\left(\begin{array}{llll}
0\ &\ 0 &\ 0 &\ v \\
0\ &\ 0 &\ 0 &\  0 \\
0\ &\ 0 &\ 0 &\  0 \\
-v\ &\ 0 &\ 0 &\  0 \\
\end{array}\right)\quad \text{and}\quad 
2\ell_{[a}\bar{m}_{b]}=\left(\begin{array}{llll}
0\ &\ 0 &\ v &\ 0 \\
0\ &\ 0 &\ 0 &\ 0 \\
-v\ &\ 0 &\ 0 &\  0 \\
0\ &\ 0 &\ 0 &\  0 \\
\end{array}\right),
\ee
both terms are independent of the structure function $f(u,\xi)$. Solving \eq{sec3:N12tensor}, the auxiliary scalar field $S_{12}$ is given by 
\be\label{sec3:RT-0scalar}
S_{12}=\frac{C(u,\bar{z})}{v(1+\epsilon z\bar{z})},
\ee
{\red where {\red the} function $C(u,\bar{z})$ is arbitrary.}
Following \eq{sec2:N14} and \eq{sec2:degenerate12}, the DW scalar and the electromagnetic scalar are solved by
\begin{align}
\xi^4&=\frac{C(u,\bar{z})^3\pp^3_{\bar{z}}\bar{f}}{2v^4(1+\epsilon{\red z}\bar{z})^{\red{2}}}, \\
\phi_2&=\sqrt{\frac{{C(u,\bar{z})\pp^3_{\bar{z}}\bar{f}}}{2}}
\frac{1}{v}\label{sec3:RT-1de-scalar}.
\end{align}
Clearly, the function $C(u,\bar{z})$ lets $\phi_2$ be independent of the structure function {\red $f$}. Thus, the electromagnetic field also satisfies the field equation {\red in} conformally flat spacetime. We can further check that the auxiliary scalar filed $S_{24} (=S_{12})$ satisfies \eq{sec3:0scalar-coformal} both {\red in} the curved spacetime and {\red in} conformally flat spacetime. 

It is worth noting that the single copy does not depend on the parameter $\epsilon$. It is the zero copy that decides what kind of sources of gravitational waves they are mapping, at least in {\red the} weak field limit. For example, given the same electromagnetic field in conformally flat spacetime, {\red following the map \eq{sec1:double copy}} the scalar field $S_{12}$ with $\epsilon=1$ will lead to a class of transverse gravitational waves whose source is time-like. {\red On the other hand, a} scalar field $S_{12}$ with $\epsilon=0$ will lead to another class of transverse gravitational waves whose source is null. 

So far, we only consider the time-dependent vacuum solutions. Next, we will investigate time-independent vacuum solutions by focusing on type D spacetimes.  More interpretations about the single copy and the zeroth copy will be discussed later.

\subsection{The case {\red of} vacuum type D solutions}

\subsubsection{Kerr-(A)dS black holes}

As we know, rotating black holes are believed {\red to be} the most typical astrophysical black holes in the universe. It is necessary to take the case of Kerr-(A)dS black holes as a specific example to study the double copy relation before going to the most general vacuum type D solutions.

The metric of Kerr-(A)dS black holes {\red in the Boyer-Lindquist coordinates} reads \cite{CARTER1968399,Sekiwa:2006qj, Cardoso:2006wa}
\begin{align}
\mm{d}s^2=\frac{\mc{R}}{\rho^2}(\mm{d}t-\frac{a}{\Sigma}\sin^2\theta \mm{d}\phi)^2&-\frac{\rho^2}{\mc{R}}\mm{d}r^2-\frac{\rho^2}{\Theta}\mm{d}\theta^2\\
&-\frac{\Theta}{\rho^2}\sin^2\theta(a \mm{d}t-\frac{r^2+a^2}{\Sigma}\mm{d}\phi)^2,
\end{align}
where
\ba
\mc{R}=(r^2+a^2)(1+\frac{r^2}{l^2})-2Mr,\quad \Theta=1-\frac{a^2}{l^2}\cos^2\theta, \\
\Sigma=1-\frac{a^2}{l^2},\quad \rho^2=r^2+a^2\cos^2\theta,\quad {\red l^2=-\frac{3}{\Lambda}},
\ea
with mass $M/\Sigma^2$ and angular momentum $J=aM/\Sigma^2$. Clearly, $M$ and $a$ can be regarded as mass parameter and angular momentum parameter, respectively.

Since the metric has already been written in the orthogonal tetrad $\{e^i\}$ $(i=1,2,3,4)$ such that $ds^2=(e^1)^2-(e^2)^2-(e^3)^2-(e^4)^2$, the null tetrad $\{e'^i\}$ then is easily given under the transformation 
\be
\begin{aligned}
e'^1=&\frac{1}{\sqrt{2}}(e^1+e^2), \quad &e'^2=\frac{1}{\sqrt{2}}(e^1-e^2),\\
e'^3=&\frac{1}{\sqrt{2}}(e^3+ie^4), \quad &e'^4=\frac{1}{\sqrt{2}}(e^3-ie^4).
\end{aligned}
\ee
Thus we have 
\ba
e'^1=\ell=\frac{1}{\sqrt{2}}\left(\sqrt{\frac{\mc{R}}{\rho^2}}\mm{d}t+\sqrt{\frac{\rho^2}{\mc{R}}}\mm{d}r-\sqrt{\frac{\mc{R}}{\rho^2}}\frac{a}{\Sigma}\mm{d}\phi\right),\\
e'^2=n=\frac{1}{\sqrt{2}}\left(\sqrt{\frac{\mc{R}}{\rho^2}}\mm{d}t+\sqrt{\frac{\rho^2}{\mc{R}}}\mm{d}r-\sqrt{\frac{\mc{R}}{\rho^2}}\frac{a}{\Sigma}\mm{d}\phi\right),\\
e'^3=m=\frac{1}{\sqrt{2}}\left(\sqrt{\frac{\Theta}{\rho^2}a}\sin\theta \mm{d}t+i \sqrt{\frac{\rho^2}{\Theta}}\mm{d}\theta-\sqrt{\frac{\Theta}{\rho^2}}\frac{(r^2+a^2)}{\Sigma}\sin\theta \mm{d}\phi\right),\\
e'^4=\bar{m}=\frac{1}{\sqrt{2}}\left(\sqrt{\frac{\Theta}{\rho^2}a}\sin\theta \mm{d} t-i \sqrt{\frac{\rho^2}{\Theta}}\mm{d}\theta-\sqrt{\frac{\Theta}{\rho^2}}\frac{(r^2+a^2)}{\Sigma}\sin\theta \mm{d}\phi\right).
\ea
We obtain the Weyl scalar
\be
\Psi_2=6\psi_2=\frac{6M}{(r+i a \cos\theta)^3},
\ee
{\red and the spin coefficients
\be\begin{aligned}
\rho^*=\mu^*=-\frac{i}{\sqrt{2}(a \cos\theta-i r)}\sqrt{\frac{\mc{R}}{\rho^2}},\\
\tau^*=\pi^*=-\frac{i a \sin\theta}{\sqrt{2}(a \cos\theta-ir)}\sqrt{\frac{\Theta}{\rho^2}},\\
\epsilon^*=\gamma^*=-\frac{a \cos\theta [l^2(r-M)+r(a^2+2r^2)]+i(a^2l^2-r^4-l^2Mr)}{2\sqrt{2}l^2(a\cos\theta-ir)}\frac{1}{\sqrt{\rho^2\mc{R}}},\\
\alpha^*=\beta^*=\frac{r\cos\theta(a^2\cos 2\theta-l^2)+ia(a^2\cos^4\theta-l^2)}{2\sqrt{2}l^2\sin\theta(a\cos\theta-ir)}\frac{1}{\sqrt{\rho^2\Theta}}.
\end{aligned}
\ee
}According to \eq{sec2:D14}, to identify the DW scalar we need to solve the auxiliary scalar field $S_{14}$. Using \eq{sec2:DS14with4eqs} {\red and the identity
\be\label{sec3-identity}
\arctan (z)=-\frac{i}{2}\log (\frac{i-z}{i+z}) \quad \quad z\in \mathbb{C},
\ee}it is not hard to obtain that
\be
S_{14}={\red \mc{K}}_1\frac{\csc^2\theta (r+i a \cos\theta)}{{\red l^4\mc{R}\Theta}},
\ee
where all of the constant coefficients have been absorbed by a constant of integration {\red $\mc{K}_1$}. 
The DW scalar {\red can then} be solved by
\be
\xi^2=\sqrt{\Psi_2 S_{14}}=\frac{\sqrt{6{\red \mc{K}}_1 M}\csc\theta}{{\red (r+i a \cos\theta)l^2\sqrt{\mc{R}\Theta}}}.
\ee
Note this is also the coefficient of DW tensor in the null tetrad \cite{Han:2022ubu}. {\red W}e are not going to talk about DW tensor in more detail in this paper, the only reason we show this is to construct {\red an} electromagnetic scalar. With the help of {\red the} auxiliary scalar $S^{(1)}_{12}$, which is solved from \eq{sec2:DS112}
\be
S^{(1)}_{12}={\red \mc{K}}_2\frac{ \csc\theta(r+i a \cos\theta)}{{\red l^2\sqrt{\mc{R}\Theta}}},
\ee
we obtain
\be
\phi_1=\frac{\xi^2}{S^{(1)}_{12}}=\frac{\sqrt{6{\red \mc{K}}_1M}}{{\red \mc{K}}_2}\frac{{\red 1}}{(r+ia\cos\theta)^2}\sim\frac{1}{(r+ia\cos\theta)^2}{\red ,}
\ee
{\red where ${\red \mc{K}}_2$, similar to ${\red \mc{K}}_1$, is an arbitrary constant of integration.} One can observe that the mass parameter $M$ is absorbed by a constant of integration. So {\red $\phi_1$} actually does not depend on the source. In addition, it is easy to verify that $(l_{[a}n_{b]}+\bar{m}_{[a}m_{b]})$ and $(l_{[a}n_{b]}+m_{[a}\bar{m}_{b]})$ are all independent of the mass parameter $M$. Hence, from \eq{sec2:nondege-tensor}, we conclude that the non-degenerate electromagnetic field we construct is independent of the source. It even satisfies the conformally invariant field equation {\red in} conformally flat spacetime, where we just need to let $M=0$ in the metric. What about the auxiliary scalar field $S^{(1,1)}_{24}$ associated to this electromagnetic field? Using the formula \eq{sec1:double copy}, one observes that
\be\label{sec3:KerrS24}
S^{(1,1)}_{24}=\frac{{\red (\phi_1)^2}}{\Psi_2}=\frac{{\red \mc{K}}_1}{({\red \mc{K}}_2)^2}\frac{1}{(r+ia\cos\theta)}\sim\frac{1}{r+ia\cos\theta}.
\ee
As we expect, this satisfies the conformally invariant scalar field equation
\be
\td{\nabla}^a\td{\nabla}_aS^{(1,1)}_{24}-\frac{1}{6}\td{R} S^{(1,1)}_{24}=0.
\ee
When $\Lambda\to0$, it describes the wave equation on Minkowski background. Thus, we have shown that the single copy and the zeroth copy of Kerr-AdS spacetimes satisfy their conformally invariant field equations {\red in} conformally flat spacetime.

Moreover, in analogy to \eq{sec3:N12tensor} of the type N case, {\red the auxiliary scalars associated with the degenerate electromagnetic fields are given by}
\be
S^{(2)}_{12}=S^{(0)}_{12}\sim \frac{1}{r+ia\cos\theta}.
\ee
Combining with \eq{sec3:KerrS24}, {\red one can see that $S^{(2)}_{12}$ and $S^{(0)}_{12}$ are equivalent to $S^{(1,1)}_{24}$ up to a constant. Therefore,} the zeroth copy connects {\red not only} gravity fields with the single copy {\red living in the conformally flat spacetime,} but also {\red DW fields with} those degenerate electromagnetic fields residing in the curved spacetime. This property is consistent with the discovery of the preceding work \cite{Han:2022ubu} in the absence of the cosmological constant $\Lambda$.

\subsubsection{The most general vacuum type D solutions}
Now, we shall investigate the Weyl double copy relation for the most general vacuum type D solutions with a cosmological constant. The metric has been given by Plebanski and Demianski\footnote{\red By doing a coordinate transformation $q\to-1/q$ and some rescalings following in eq. (3) of Ref. \cite{Griffiths:2005qp}, we will go back to the modified form of the metric applied in the preceding work \cite{Han:2022ubu}.} \cite{Plebanski:1976gy},
\ba
ds^2=\frac{1}{(p+q)^2}\left(-\frac{1+(pq)^2}{\mc{P}}\mm{d}p^2-\frac{\mc{P}}{1+(pq)^2}(\mm{d}\sigma+q^2\mm{d}\tau)^2-\right.\\
\left. \frac{1+(pq)^2}{\mc{L}}\mm{d}q^2+\frac{\mc{L}}{1+(pq)^2}(-p^2\mm{d}\sigma+\mm{d}\tau)^2\right),
\ea
where the structure functions read
\ba
\mc{P}=(-\frac{\Lambda}{6}+\gamma)+2np-\epsilon p^2+2mp^3+(-\frac{\Lambda}{6}-\gamma) p^4,\\
\mc{L}=(-\frac{\Lambda}{6}-\gamma)+2nq+\epsilon q^2+2mq^3+(-\frac{\Lambda}{6}+\gamma) q^4,
\ea
$m$ and $n$ are dynamical parameters measuring the curvature, $\epsilon$ and $\gamma$ are called kinematical parameters which will affect the properties of the solutions. 

By choosing null tetrad 
\ba
\ell=\frac{1}{{\red \sqrt{2}}(p+q)}\left(\sqrt{\frac{1+(pq)^2}{\mc{L}}}dq-p^2\sqrt{\frac{\mc{L}}{1+(pq)^2}}d\sigma+\sqrt{\frac{\mc{L}}{1+(pq)^2}}d\tau\right),\\
n=\frac{1}{{\red \sqrt{2}}(p+q)}\left(-\sqrt{\frac{1+(pq)^2}{\mc{L}}}dq-p^2\sqrt{\frac{\mc{L}}{1+(pq)^2}}d\sigma+\sqrt{\frac{\mc{L}}{1+(pq)^2}}d\tau\right),\\
m=\frac{1}{{\red \sqrt{2}}(p+q)}\left(\sqrt{\frac{1+(pq)^2}{\mc{P}}}dp+i \sqrt{\frac{\mc{P}}{1+(pq)^2}}d\sigma+i q^2\sqrt{\frac{\mc{P}}{1+(pq)^2}}d\tau\right),\\
\bar{m}=\frac{1}{{\red \sqrt{2}}(p+q)}\left(\sqrt{\frac{1+(pq)^2}{\mc{P}}}dp-i \sqrt{\frac{\mc{P}}{1+(pq)^2}}d\sigma-i q^2\sqrt{\frac{\mc{P}}{1+(pq)^2}}d\tau\right),
\ea
we obtain
\be
\Psi_2=6\psi_2={\red 6}(m+i n)\left(\frac{p+q}{1-ipq}\right)^3.
\ee
Obviously, the cosmological constant does not affect the Weyl scalar according to the above result. 
{\red Some spin coefficients are given by
\be
\begin{aligned}
\rho^*=\mu^*=\frac{(p^2-i)\sqrt{\mc{L}(q)}}{\sqrt{2}(pq+i)\sqrt{1+p^2q^2}},\\
\tau^*=-\pi^*=\frac{(q^2-i)\sqrt{\mc{P}(p)}}{\sqrt{2}(pq+i)\sqrt{1+p^2q^2}},\\
\epsilon^*=\gamma^*=\frac{2(p^2+2pq+i)\mc{L}(q)-(p+q)(pq+i)\mc{L}'(q)}{4\sqrt{2}(pq+i)\sqrt{1+p^2q^2}\sqrt{\mc{L}(q)}},\\
\beta^*=-\alpha^*=\frac{2(q^2+2pq+i)\mc{P}(p)-(p+q)(pq+i)\mc{P}'(q)}{4\sqrt{2}(pq+i)\sqrt{(1+p^2q^2)\mc{P}(p)}}.
\end{aligned}
\ee}Solving \eq{sec2:DS14with4eqs} {\red with the help of \eq{sec3-identity}}, the auxiliary scalar field $S_{14}$ is given by
\be
S_{14}={\red \mc{D}}_1\frac{(p+q)^3(1-i p q)}{\mc{P}(p)\mc{L}(q)},
\ee 
where ${\red \mc{D}}_1$ is a constant of integration.
Then from \eq{sec2:D14} we have
\be
\xi^2=\frac{\sqrt{{\red 6}{\red \mc{D}}_1(m+i n)}}{\sqrt{\mc{P}(p)\mc{L}(q)}}\frac{(p+q)^3}{1-ipq}.
\ee
Recalling \eq{sec2:nondege-tensor}, one observes that
\be
2\left(\ell_{[a}n_{b]}+\bar{m}_{[a}m_{b]}\right)=\left(\begin{array}{llll}
0\ &\ 0 &\ i A & i p^2 A \\
0\ &\ 0 &\ q^2 A &\  -A \\
-i A\ &\-q^2 A&\ 0 &\  0 \\
-i p^2 A\ &\ A &\ 0 &\  0 \\
\end{array}\right),
\ee
and
\be
2\left(\ell_{[a}n_{b]}+\bar{m}_{[a}m_{b]}\right)=\left(\begin{array}{llll}
0\ &\ 0 &\ -i A & -i p^2 A \\
0\ &\ 0 &\ q^2 A &\  -A \\
i A\ &\-q^2 A&\ 0 &\  0 \\
i p^2 A\ &\ A &\ 0 &\  0 \\
\end{array}\right),
\ee
where $A=\frac{(p+q)^2}{1+p^2q^2}$. Clearly, they are independent of the dynamical parameters. {\red Consequently,} if the electromagnetic scalar is independent of the {\red deviation-information}, the electromagnetic field we construct will be independent of the {\red deviation-information as well} and the field equation will hold even {\red in} conformally flat spacetime. From \eq{sec2:DS112}, one obtains
\be
S^{(1)}_{12}={\red \mc{D}}_2\frac{(p+q)(1-ipq)}{\sqrt{\mc{P}(p)\mc{L}(q)}},
\ee
where ${\red \mc{D}}_2$ is a constant of integration.
Following \eq{sec2:nondegenerate12}, the non-degenerate electromagnetic scalar $\phi_1$ is given by
\be
\phi_1=\sqrt{\frac{{\red 6}{\red \mc{D}}_1(m+in)}{{\red \mc{D}}_2}}\frac{(p+q)^2}{(1-ipq)^2}\sim\frac{(p+q)^2}{(1-ipq)^2}.
\ee
One can see that the dynamical parameters which measure the curvature are absorbed by the constants of integration, so $\phi_1$ is independent of the dynamical parameters. Thus, we have discovered a {\red particular} non-degenerate electromagnetic field {\red which is independent of the deviation-information}. Correspondingly, the auxiliary scalar field $S^{({\red 1,1})}_{24}$ is given by
\be
S^{(1,1)}_{24}=\frac{(\phi_1)^2}{\Psi_2}=\frac{{\red \mc{D}}_1}{({\red \mc{D}}_2)^2}\frac{p+q}{1-ipq}\sim\frac{p+q}{1-ipq}.
\ee
It is easy to check that $S_{24}^{({\red 1,1})}$ satisfies the conformal invariant {\red field} equation \eq{sec3:0scalar-coformal} {\red in} conformally flat spacetime, where we just need to set $m=n=0$. 

Therefore, for vacuum type D solutions with a cosmological constant, the single copy and the zeroth cop{\red ies} satisfy their conformal invariant field equations {\red in} conformally flat spacetime. When $\Lambda\to0$, the background goes back to Minkowski spacetime {\red and} the situation is consistent with the previous result \cite{Luna:2018dpt}.

In addition, similar to the case of Kerr-AdS spacetime, for the general vacuum type D solutions with or without a cosmological constant, we find that
\be
S^{(0)}_{12}=S^{(2)}_{12}\sim\frac{p+q}{1-i p q}\sim S^{(1,1)}_{24}.
\ee
{\red Thus, not only does the zeroth copy connect gravity fields to the single copy, but it also links DW fields to degenerate electromagnetic fields living in curved spacetime.} Recalling the previous section, {\red it is evident that} this property also {\red applies to} non-twisting type N solutions. While, {\red distinct} from the type N cases, the zeroth copy now does not possess any extra information about the source. This is {\red mirrored clearly} by the double copy scalar relation $(\Psi_2)^{1/3}=(\phi_2)^{1/2}=S^{(1,1)}_{24}$. Therefore, we find that only for the time-dependent solutions, the zeroth copy carries extra information about the source. This provides support for constructing other exact time-dependent radiation solutions {\red in} future work.

\section{Discussion and Conclusions}
\label{sec4}
In this paper, {\red using} DW spinors (massless spin-$1/2$ spinors) as basic units, we constructed {\red a particular set of} electromagnetic fields in {\red 4-dimensional} non-twisting vacuum type N and vacuum type D spacetimes {\red in} the presence of a cosmological constant {\red $\Lambda$}. {\red These electromagnetic fields are independent of the deviation-information for a given curved metric. Thus they also satisfy the field equation in conformally flat spacetime}. {\red R}egarding {\red these electromagnetic} fields as the single copies {\red in the curved (A)dS spacetime}, we verified the Weyl double copy {prescription}. {\red W}e found that the single and zeroth copies satisfy {\red the} conformally invariant field equations {\red in} conformally flat spacetime both for the time-dependent solutions (type N cases) and time-independent solutions (type D cases). When $\Lambda\to0$, the result reduces to the original case. Namely, they satisfy {\red the} field equations {\red in} Minkowski spacetime. {\red This is an intriguing outcome.}  {\red For Kerr-Schild ($\Lambda$) double copy prescription \cite{Carrillo-Gonzalez:2017iyj}, the single and zeroth copies satisfy different equations for time-independent solutions and time-dependent solutions.  Specifically, in time-independent cases, the zeroth and single copies satisfy the conformally invariant field equations in conformally flat spacetime; whereas, in time-dependent cases, the zeroth copy satisfies the wave equation and does not admit good conformal transformation properties anymore. Moreover, the single copy does not satisfy the conformally invariant Maxwell's field equation because of an extra term proportional to the Ricci scalar appearing in the equation of motion. Therefore,} from this point of view, the Weyl double copy {\red prescription} appears as a more fundamental map between gravity theory and gauge theory. {\red This is also consistent with the fact that the Kerr-Schild double copy prescription is linear, whereas the Weyl double copy prescription is essentially more general.}  

{\red Apart from the above results}, we found that the preceding finding \cite{Han:2022ubu} also holds in the presence of $\Lambda$. {\red Not only does} the zeroth copy connect gravity fields with the single copy {\red in the conformally flat spacetime, but it} also connects {\red DW fields with} degenerate electromagnetic fields in the curved spacetime, both for non-twisting vacuum type N solutions and vacuum type D solutions. More interestingly, we found that the zeroth copy plays {\red a more} important role {\red than expected} for time-dependent radiation solutions (type N cases). Unlike the single copy, which only restricts the form of the structure function, the zeroth copy carries {\red additional} information {\red characterizing the} curved spacetimes {\red it is} mapping. {\red Specifially} for the Robinson-Trautman ($\Lambda$) class, we {\red discovered that} it is the zeroth copy that {\red determines} whether the sources of associated gravitational waves are time-like, null, or {\red space-like}, {\red at least in the weak field limit}. {\red This result is reminiscent of previous research on the fluid/gravity duality \cite{Keeler:2020rcv}, which showed that all of the information about the fluid is encoded in the zeroth copy for the type N case.} {\red Their work further supports our result that the zeroth copy can indeed carry additional information compared with the single copy. We hope this discovery will contribute to} constructing other exact time-dependent radiation solutions.

All in all, we have shown {\red explicitly} that the single copy and the zeroth copy satisfy conformally invariant equations {\red in} conformally flat spacetime by {\red concentrating} on non-twisting vacuum type N and vacuum type D solutions. {\red Several novel interpretations of the Weyl double copy prescription are provided, particularly with regard to} the zeroth copy. Next, it would be intriguing to check whether the generalized Weyl double copy holds asymptotically for the algebraically general case with a cosmological constant. {\red I}t is {\red also }significant to investigate the applications of the Weyl double copy on astrophysical observations, such as the specific correspondence between the source of gravitational waves and the Weyl double copy.  {\red In addition, a natural progression of this work is to analyse the classical Weyl double copy in the flat spacetime instead of the (A)dS background. That would be essential for establishing a bridge between gravity theory and gauge theory. The cosmological constant, in this case, should be considered as the source of the single and zeroth copies. Further studies which treat the Minkovski spacetime as the background of the Weyl double copy prescription in the presence of a cosmological constant will need to be undertaken in the future. }Since all of the discussion in this paper is limited to $4$-dimensional spacetimes, it {\red would also be worthwhile to extend} the study to high-dimensional spacetimes.

\section*{Acknowledgements}
The author would like to thank Niels A. Obers for useful comments on the manuscript. The author thanks the Theoretical Particle Physics and Cosmology section at the Niels Bohr Institute for support. This work is also financially supported by the China Scholarship Council.

\bibliographystyle{JHEP}
\bibliography{refss}

\end{document}

%% file: defs.tex
\newcommand{\red}{\color{red}}

